\documentclass[aps,prl, twocolumn,nofootinbib,10pt]{revtex4}
\usepackage{epsfig}
\newcounter{fig}   \newcommand{\lbfig}[1]{\refstepcounter{fig}
\label{#1} }
\usepackage{bm}

\usepackage{amsmath}
\usepackage{amsfonts}
\usepackage{graphicx}
\usepackage{amssymb}
\usepackage{amssymb}
\usepackage{latexsym}
\usepackage{hyperref}
\usepackage{array}
\usepackage{bbold}

\newcommand{\bea}{\begin{eqnarray}}
\newcommand{\eea}{\end{eqnarray}}
\newcommand{\be}{\begin{equation}}
\newcommand{\ee}{\end{equation}}
\newcommand{\re}[1]{(\ref{#1})}

\usepackage{xcolor}

\begin{document}
\title{Kinks bounded by fermions}
\begin{abstract}

We present and study new mechanism of interaction between the solitons based on the exchange interaction mediated
by the localized fermion states. As particular examples, we consider solutions of simple 1+1 dimensional scalar
field theories with self-interaction potentials, including sine-Gordon model and the polynomial $\phi^4$,
$\phi^6$ models, coupled to the Dirac fermions with back-reaction. We discover that there is an additional
fermion exchange interaction between the solitons, it leads to the formation of static multi-soliton bound
states. Further, we argue that similar mechanisms of formation of stable coupled multi-soliton configurations can
be observed for a wide class of physical systems.

\end{abstract}
\author{Ilya Perapechka}
\affiliation{Department of Theoretical Physics and Astrophysics,BSU, Minsk 220004, Belarus}
\author{Yakov Shnir}
\affiliation{BLTP, JINR,\\Dubna 141980, Moscow Region, Russia,\\
}

\maketitle
\section{Introduction}

One of the most interesting features of the topological solitons, like kinks, vortices or monopoles, see, e.g.
\cite{Manton:2004tk,Shnir2018}, is the remarkable relation between the topological charge of the configuration
and the number of fermionic zero modes localized on a soliton. The fundamental Atiyah-Patodi-Singer index theorem
requires one normalizable fermionic zero mode per unit topological charge \cite{APS}. Moreover, apart zero modes,
most configurations support existence of a tower of localized fermionic modes with non-zero energy.

The fermion zero modes of the solitons have been studied for many decades, these states localized  on the vortices
were discussed first in \cite{zeromode}. There has been substantial
interest in study of these localized states in various dimensions, examples of such are fermion modes of the
kinks \cite{Dashen:1974cj,Jackiw:1975fn,Chu:2007xh,Liu:2008pi}, monopoles
\cite{Rubakov:1982fp,Callan:1982au},
sphalerons \cite{Nohl:1975jg,Boguta:1985ut} and
skyrmions \cite{Hiller:1986ry,Kahana:1984dx,Kahana:1984be,Ripka:1985am}.
Existence of localized fermions leads to many interesting and unusual phenomena such as
fermion number fractionization \cite{Jackiw:1975fn,Jackiw:1981ee},
monopole catalysis \cite{Rubakov:1982fp,Callan:1982au}, and string superconductivity
in cosmology \cite{Witten:1984eb}, or in chiral superconductors \cite{Volovik}.

Fermion zero modes naturally appear in supersymmetric theories,  several simple examples are
$N=1$ chiral scalar superfield in 1+1 dimesions \cite{DiVecchia:1977nxl},
or supersymmetric extensions of the $\mbox{O}(3)$ non-linear sigma model
\cite{Novikov:1984ac} and baby-Skyrme model \cite{Adam:2011hj,Queiruga:2016jqu}, in such a case
fermion zero mode is generated via  supersymmetry transformations of the boson field of the static soliton.
The breaking of supersymmetry of the configurations, in agreement with the index theorem
shows a spectral flow of the eigenvalues of the Dirac operator with some number of normalizable bounded modes
crossing zero.

The typical assumption in most of such considerations is that the
the spinor field does not backreact on the soliton
\cite{Dashen:1974cj,Jackiw:1975fn,Chu:2007xh,Gibbons:2006ge},  moreover,
only the fermion zero modes were considered in most cases.
A completely different approach to the problem was proposed in our previous works
\cite{Perapechka:2018yux,Perapechka:2019dvc,Klimashonok:2019iya} where we reconsidered this problem consistently.
We found that the back-reaction of the localized fermions significantly modifies both the spectral flow and
the structure of the coupled configurations. In particular, it results in deformations of the solitons, which
to a certain extent, resemble the excitations of the internal modes of the scalar configurations
\cite{Klimashonok:2019iya,Manton:1996ex,Manton:2019xiq}.

In contrast to previous studies, our focus here is not to consider localized fermion states but rather
investigate the pattern of interactions between the solitons in the presence of the additional exchange
interaction mediated by localized fermions with finite energy. Our aim  is to study the collective fermion modes
in the system of kinks in 1+1 dimensional scalar field theories, including sine-Gordon model and the polynomial
$\phi^4$, $\phi^6$ models, coupled to the Dirac fermions with back-reaction. Notably, solitons in all these
models may support fermion zero mode, which does not affect the kink for any values of the Yukawa coupling.
Further, a single $\phi^4$ kink, or a kink in the sine-Gordon model
\cite{Brihaye:2008am} may bound other localized fermion modes with finite energy, the number of these modes
extracted from the positive and negative continuum increases as the Yukawa coupling becomes stronger.
The novel aspect is that exponentially localized fermion bound state may appear in the system of kinks,
just in the same way as
the collective scalar bound states become trapped by the kink-antikink potential in the $\phi^6$ model
\cite{Dorey:2011yw}.

In this Letter we investigate  multi-soliton--fermion systems with back-reaction
numerically and elucidate the mechanism for appearance of the collective fermionic modes.
Our computations reveal sequences of new bounded collective fermion states in various
systems of solitons. Further, we show that these localized modes give rise to
additional interaction between the solitons.

{\it The model.~}
The 1+1 dimensional field theory we are interested in
is defined by the following Lagrangian
\begin{equation}
\mathcal{L}=\frac{1}{2}\partial_\mu\phi\partial^\mu \phi
+ \bar \psi\left[ i\gamma^\mu \partial_\mu  -m - g\phi \right]\psi
-U(\phi) \, ,
\label{lag}
\end{equation}
where $U(\phi)$ is a potential of the self-interacting real scalar field $\phi$,
$\psi$ is a two-component Dirac spinor and $m,g$ are the bare
mass of the fermions and the Yukawa coupling constant,
respectively\footnote{Here we are using rescaled dimensionless variables.}.
The matrices $\gamma_\mu$ are
$\gamma_0=\sigma_1$, $\gamma_1=i\sigma_3$ where $\sigma_i$ are the Pauli matrices, and
$\bar \psi = \psi^\dagger\gamma^0$.
The sine-Gordon (sG) model corresponds to the periodic potential $U(\phi)=1-\cos \phi$
with infinite number of degenerate vacua $\phi_0=2\pi n$, $n\in \mathbb{Z}$,
the $\phi^4$ model corresponds to the quartic potential $U(\phi)=\frac12 \left(1-\phi^2\right)^2$
with two vacua $\phi_0\in \left\{-1,1 \right\}$, and the $\phi^6$ theory is defined by
$U(\phi)=\frac12 \phi^2\left(1-\phi^2\right)^2$
with triple degenerated vacuum $\phi_0\in \left\{-1,0,1 \right\}$.

Using the standard parametrization for a two-component spinor
\vskip -2em
$$
\label{ansFer}
\psi= e^{-i\epsilon t}\left(
\begin{array}{c}
u(x)\\
v(x)
\end{array}
\right)\, ,
$$
we obtain the following coupled system of dynamical equations
\vskip -2em
\be
\begin{split}
\phi_{xx} + 2g uv -U^\prime &=0\, ;\\
u_x+(m+g\phi)u&=\epsilon v\, ;\\
-v_x+(m+g\phi)v&=\epsilon u\, .
\end{split}
\label{eq2}
\ee
This system is supplemented by the normalization condition $\int\limits_{-\infty}^\infty\! dx\, (u^2+v^2)\!=\!1$ which we impose
as a constraint on the system \re{eq2}.

For all previously mentioned potentials, in the decoupled limit $g=0$,
the  model \re{lag} supports spatially localized static topological solitons, the kinks:
\begin{equation}
 \Phi_{sG}=4 \arctan e^{x},\, \Phi_{{\small \phi^4}}=\tanh (x), \,
 \Phi_{{\small \phi^6}}=\pm\sqrt{\frac{1\!+\!\tanh x}{2}}.
 \label{kinks}
\end{equation}
The spectrum of linearized perturbations of the kinks always includes a translational zero mode $\Phi(x)\to \Phi(x-x_0)$
and the continuum modes (e.g. see \cite{Manton:1996ex,Shnir2018}).
A peculiarity of the $\phi^4$ model is that, apart these modes, there is also an internal bosonic
mode of the kink $\Phi_{{\small \phi^4}}$.

The coupled pair of the first order differential equations in \re{eq2} over the background provided by the kinks
\re{kinks} can be transformed into two decoupled
Schr\"odinger-type equations for the spinor components $u$ and $v$ \cite{Dashen:1974cj}
\be
\begin{split}
-u_{xx}+\left( (m+g \Phi)^2 - g \Phi_x \right) u&=\epsilon^2 u\, ;\\
-v_{xx}+\left( (m+g \Phi)^2 + g \Phi_x \right) v&=\epsilon^2 v\, .
\end{split}
\label{eq3}
\ee
We can easily see that the $\phi^4$ potential  supports exponentially localized
fermion zero mode $\epsilon_0=0$, up to some modifications of the coupling term \cite{Brihaye:2008am},
or the potential, such a mode exists for all types of the configurations above.
The zero eigenvalue
does not depend on the Yukawa coupling $g$, there is no level crossing spectral flow in one spatial dimension.
In the special case of the $N=1$
supersymmetric generalization of the model \re{lag} \cite{DiVecchia:1977nxl}
this mode is generated via SUSY transformation of the translational mode of the static kink.

One of essential features of the symmetric $\phi^4$ kink is that for large values of the Yukawa coupling,
other localized fermionic states with non-zero energy eigenvalues $| \varepsilon | < |g-m| $ appear
in the spectrum \cite{Dashen:1974cj,Jackiw:1975fn,Chu:2007xh,Klimashonok:2019iya,Liu:2008pi,Brihaye:2008am}.

Consideration of the fermion modes bounded to a soliton usually
invokes a simplifying assumption that the back-reaction of the
localized fermions is negligible \cite{Dashen:1974cj,Jackiw:1975fn,Chu:2007xh,Gibbons:2006ge}.
However, coupling to the higher localized modes may significantly distort the $\phi^4$ kink, its
profile deforms as a fermion occupies an energy level \cite{Klimashonok:2019iya}. Further,
since such exponentially localized fermion modes may occur in multisoliton systems, localized
fermions could mediate the exchange interaction between the solitons.
In particular,
they may appear as a collective state trapped by the kink-antikink pair \cite{Chu:2007xh,Brihaye:2008am},
in a way analogous to the appearance of the collective bosonic modes in the $\phi^6$ model \cite{Dorey:2011yw}.

{\it Numerical results.~}
To find a numerical solution of the complete system of integral-differential equation \re{eq2} with
the normalization condition on the spinor field we used 8th order finite-difference method.
The system of equations is discretized on a uniform grid with usual size of 5000 points.
For the sake of simplicity we considered the fermions with zero bare mass, $m=0$.
The emerging system of nonlinear algebraic equations is solved
using a modified Newton method. The underlying linear system is solved with the Intel MKL PARDISO sparse direct solver.
The errors are on the order of $10^{-9}$.

First, we consider $K\bar K$ pair in the $\phi^4$ model with fermions taking into account the back-reaction of the localized
modes. Note that in the decoupled limit $g=0$
the $K\bar K$ pair is not a solution of the field equations, there is an attractive interaction between the kink
and the antikink, the only solution of the $\phi^4$ model in the topologically trivial sector is the vacuum
$\phi_0=0$.
Numerical computations shows that,
as the Yukawa coupling increases slightly above zero, a non-topological soliton emerge in the scalar sector,
this lump is linked to a localized fermionic mode extracted from the positive continuum. As $g$ increases further, the
lump becomes larger, its top corresponds to the maximum of the fermion density distribution. Solutions of that type
on the background of a well-separated approximated kink-antikink pair were discussed in \cite{Chu:2007xh}.
We will refer to the modes of that type to as $L_k$-modes, where the index $k$ corresponds to the
minimal number of nodes of the components.

Taking into account the back-reaction of the fermions, we found another, lower energy branch of solutions.
It also emerges from the positive energy continuum, corresponding solutions represent well-separated
$K\bar K$ pair with two fermion modes bounded to each soliton. At finite separation, the spinor field represents
a symmetric linear combination of two localized quasi-zero modes. Further increase of the coupling yields stronger attraction
between the constituents, the fermion density distribution possess two peaks along this branch.

\begin{figure}
 \begin{center}
\includegraphics[width=0.19\textwidth, trim={60, 20, 100, 60}, clip = true]{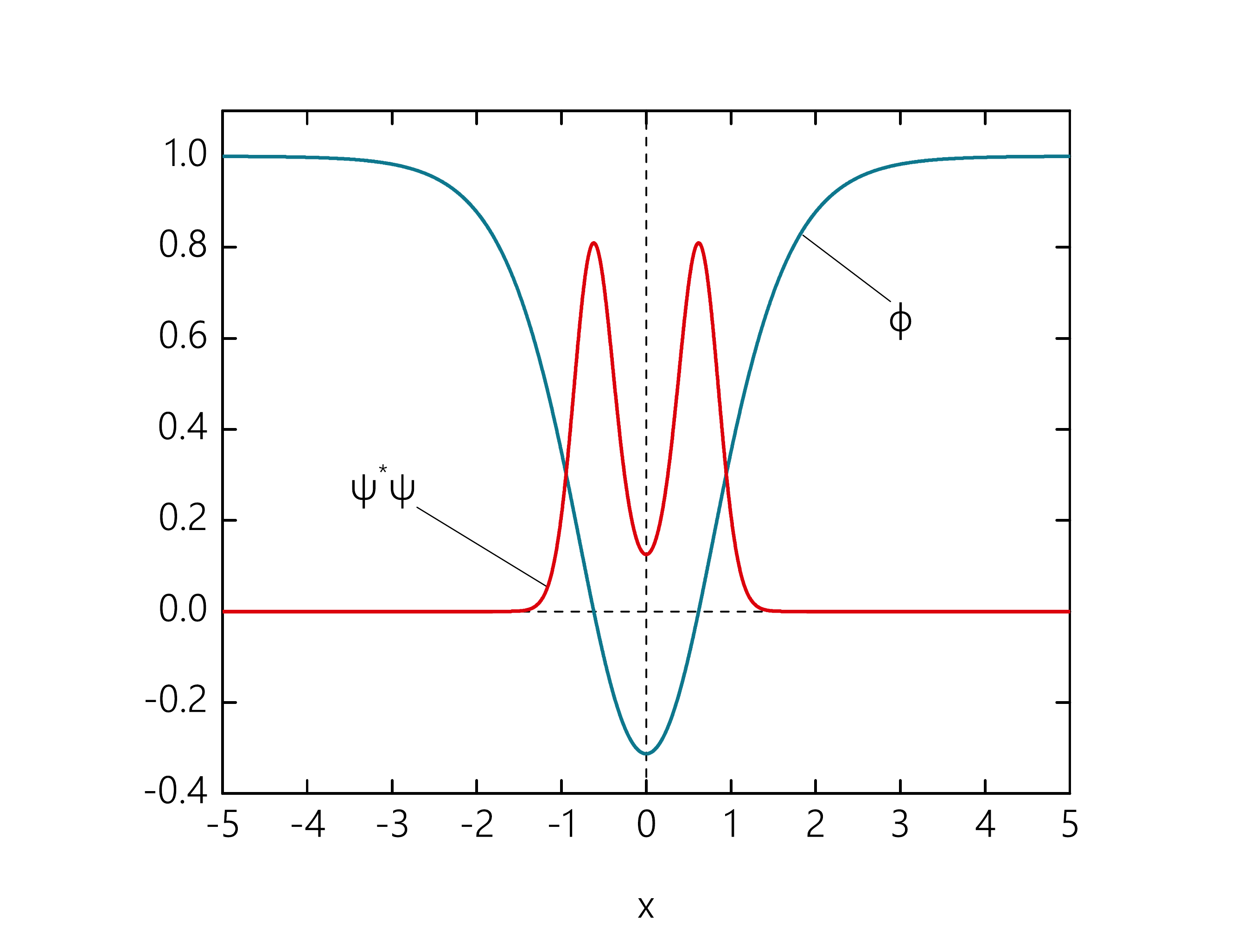}
\includegraphics[width=0.285\textwidth, trim={0, -30, 0, 0}]{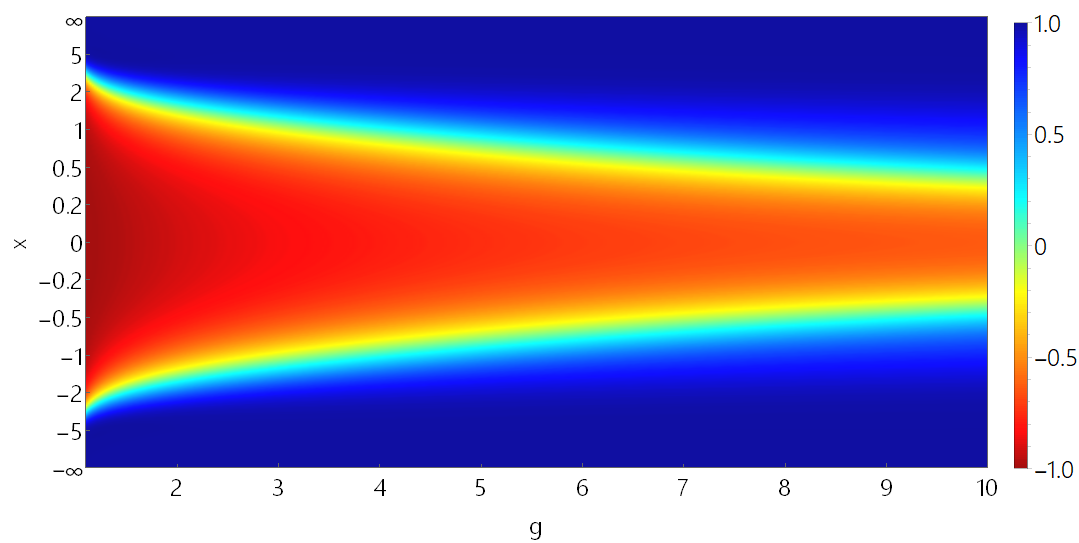}
\includegraphics[width=0.19\textwidth, trim={60, 20, 100, 60}, clip = true]{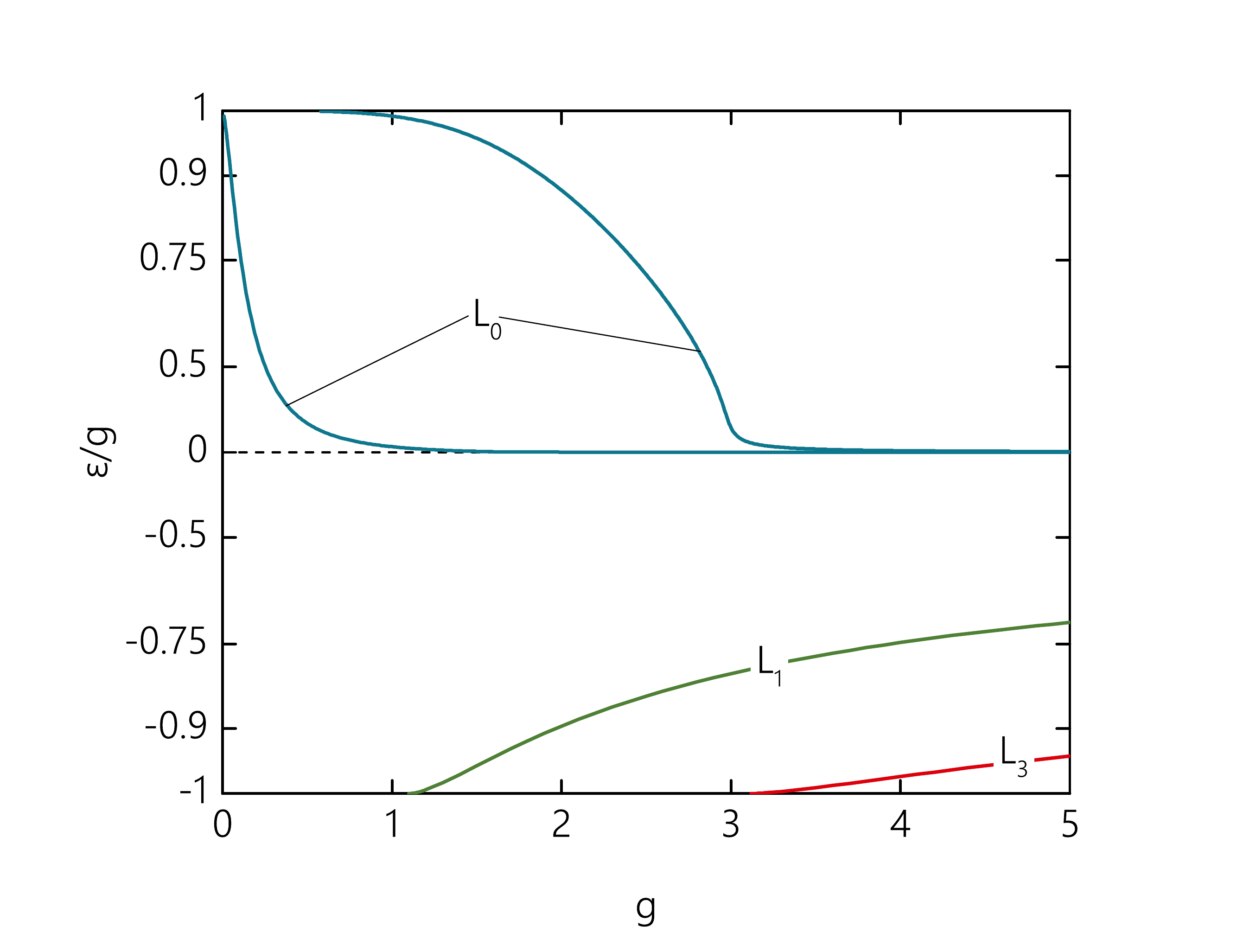}
\includegraphics[width=0.285\textwidth, trim={0, -30, 0, 0}]{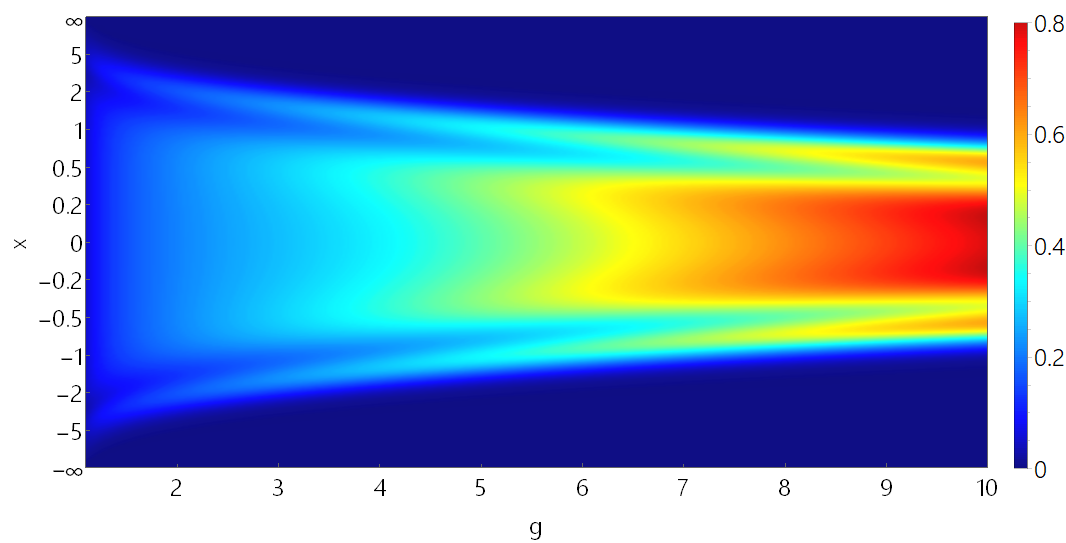}
\end{center}
 \caption{$\phi^4$ kink-antikink pair bounded by fermions. Profiles of the scalar field and
 fermion density distribution of the collective mode $L_1$ at $g=1$ (upper left),
 normalized energy  ${\epsilon}/{g}$ of the localized fermionic states
as a function of the Yukawa coupling $g$
for several modes (lower left), scalar field of the configuration bounded to the mode $L_1$
(upper right) and fermionic density
distribution of the mode $L_1$ (bottom right) vs Yukawa coupling $g$.}
\lbfig{Fig1}
\end{figure}
Further, we observe that in the range of values of the coupling $0<g<1$ there are only quasi-zero collective modes.
As we increase the Yukawa coupling further,
we found a sequence of higher collective states bounded by the $K\bar K$ pair, see Fig.~\ref{Fig1}.
Similar to the modes localized on the $\phi^4$ kink, first such mode $L_1$ appears at $g>1$, the mode
$L_3$ is extracted from the negative continuum at $g>3$, etc. Evidently, for strong coupling the back-reaction of the
fermions cannot be neglected.

Since fermion bound states exist on a single $\phi^4$ kink, all collective modes localized on the $K\bar K$ pair
represent various linear combinations of these states. On the contrary,
the model \re{lag}, for a given choice of the Yukawa coupling, do not support
fermion states bounded neither to
a single sG kink, nor to a single $\phi^6$ kink. However, the
situation can be different in the presence of two, or more solitons. Indeed, we found collective
fermions localized on various multi-kink configurations.

\begin{figure}
 \begin{center}
\includegraphics[width=0.19\textwidth, trim={49, 20, 100, 60}, clip = true]{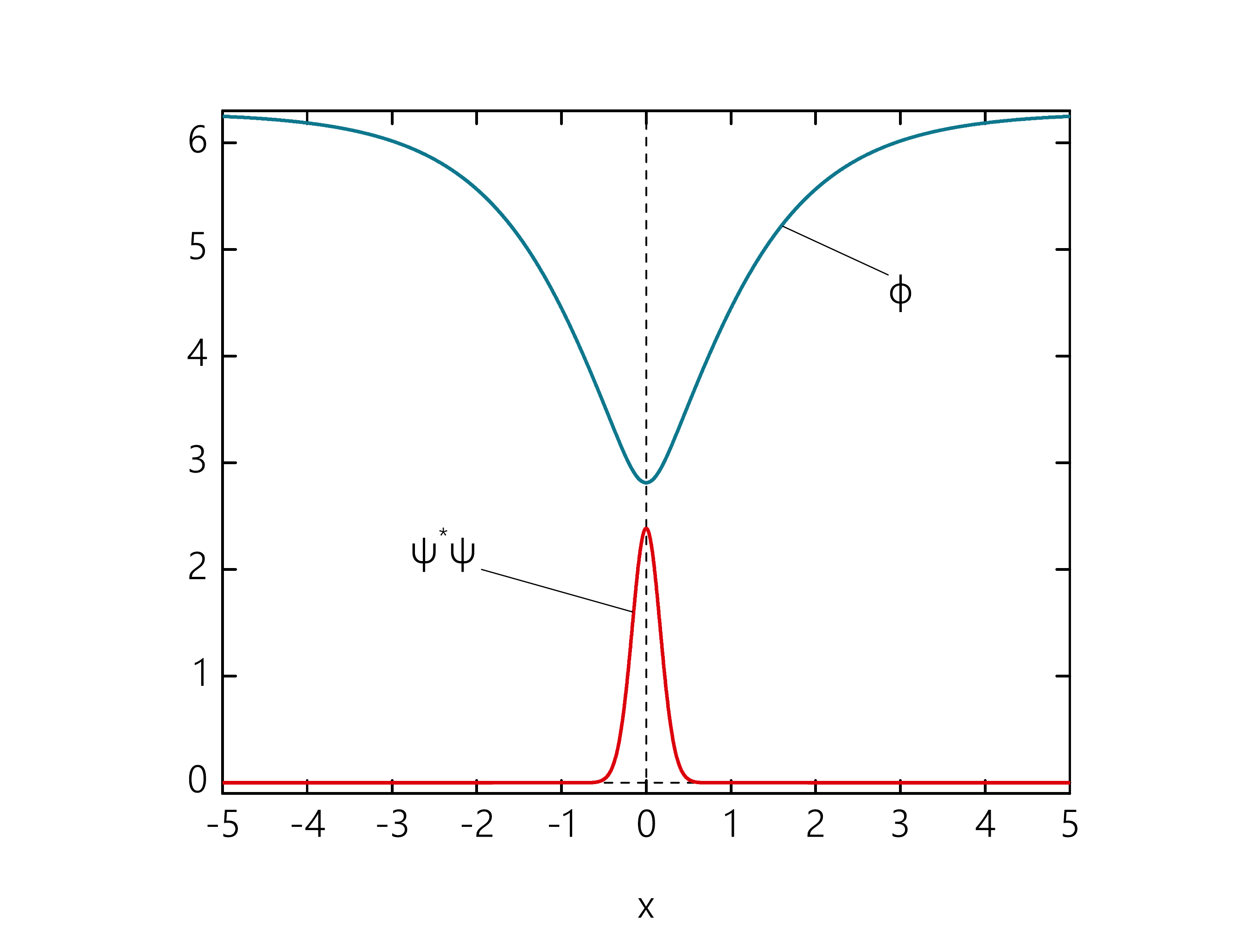}
\includegraphics[width=0.285\textwidth, trim={0, -20, 0, 0}]{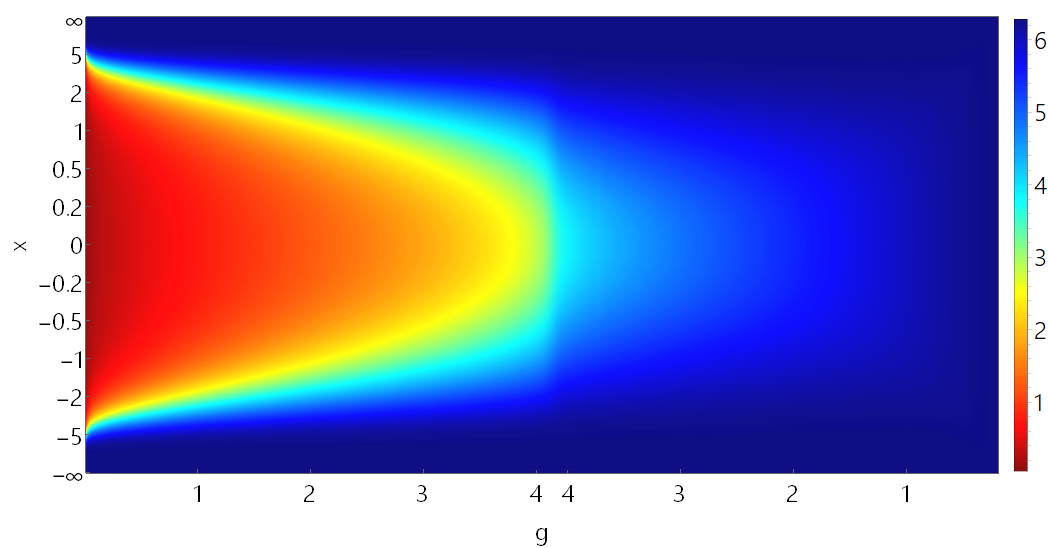}
\includegraphics[width=0.19\textwidth, trim={49, 20, 100, 60}, clip = true]{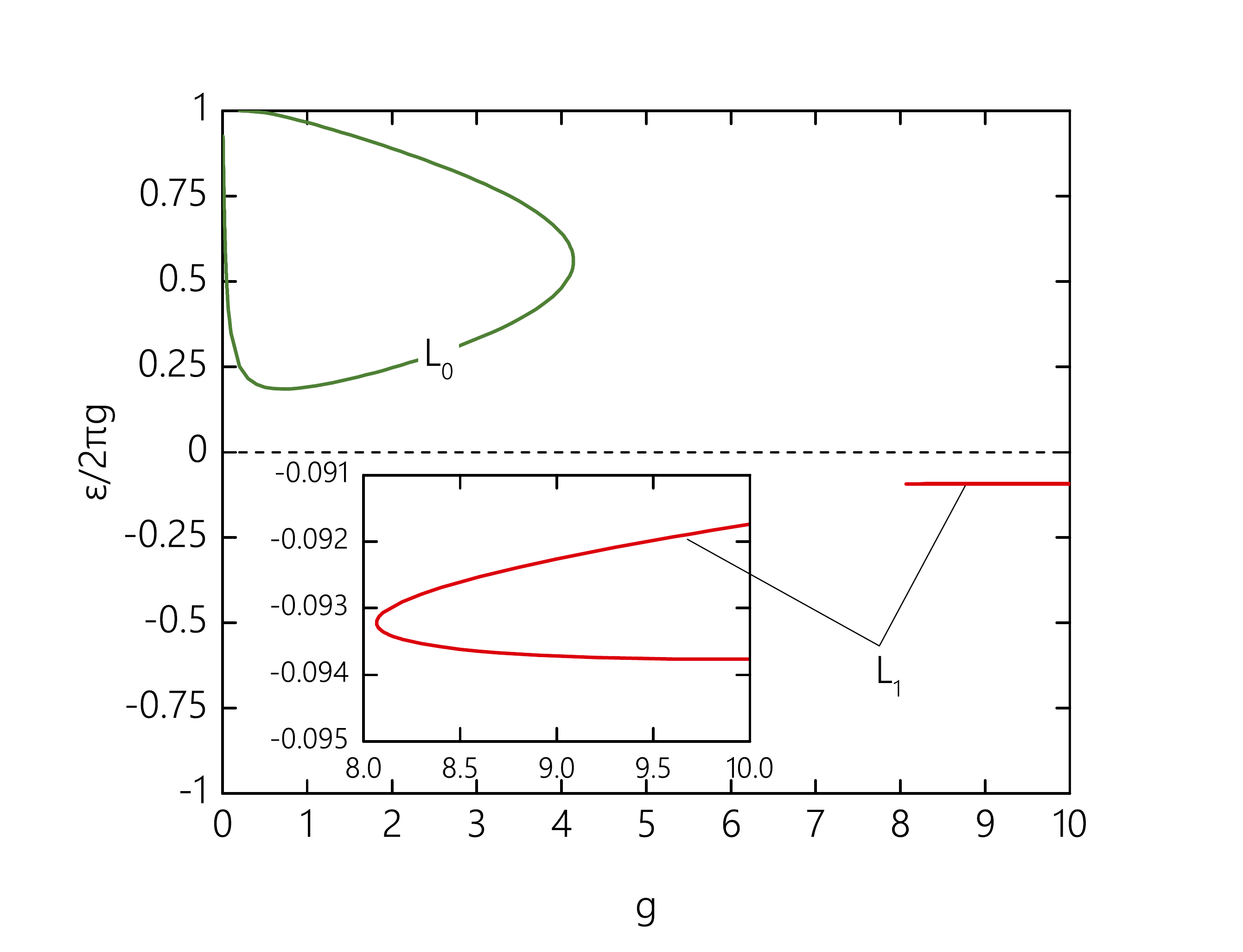}
\includegraphics[width=0.285\textwidth, trim={0, -20, 0, 0}]{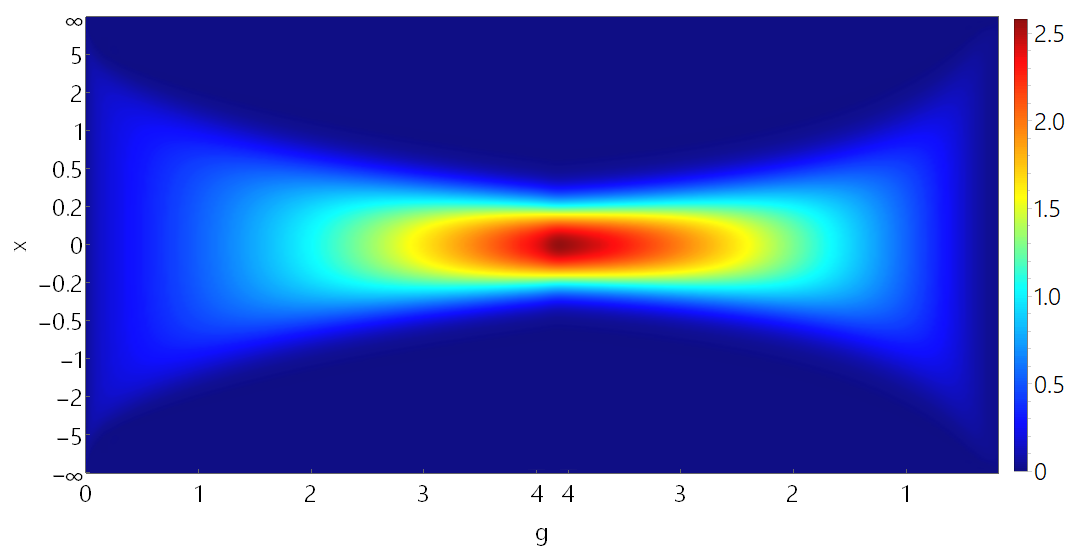}
\end{center}
 \caption{sG kink-antikink pair bounded by fermions. Profiles of the scalar field and
 fermion density distribution of the collective mode $L_0$ at $g=1$ (upper left),
 normalized energy  ${\epsilon}/{g}$ of the localized fermionic states
as a function of the Yukawa coupling $g$
for modes $L_0$ and $L_1$ (bottom left), scalar field of the configuration bounded to the mode $L_0$
(upper right) and fermionic density
distribution of the mode $L_0$ (bottom right) vs Yukawa coupling $g$.}
\lbfig{Fig2}
\end{figure}

The pattern we observed in the sG model is similar yet different from the situation above.
Again, a lump coupled to a localized nodeless fermion state $L_0$ emerges in the topologically trivial scalar sector,
as the Yukawa coupling
increases above zero, see Fig.~\ref{Fig2}. As the coupling gradually increases, the profile of the scalar field
becomes more pronounced, it approaches the shape of a "frozen" sG breather.
At some critical upper value of the Yukawa coupling this, lower in energy branch,
bifurcates with the second, higher in energy branch,
which extends all the way backwards to the limit of zero coupling, as seen in Fig.~\ref{Fig2}. The kinks separate
along this branch while a single fermion mode $L_0$ remains at the center of the configuration,
it dissolves, as the kinks become infinitely separated.

A novel feature of the sG system with backreacting fermions is that a new type of solutions $L_1$
emerge at the bifurcation point $g>8$, see Fig.~\ref{Fig2}.
The corresponding two branches are not linked to the continuum, further increase of the Yukawa
coupling along both branches yields strong distortion
of the profile of the $K\bar K$ configuration which tends toward deformed $2K - 2\bar K$ coupled system.
Physically, the branch structure emerge because of existence of two linear combinations
of the fermion modes, localized on each component.

Indeed, a tower of localized fermion modes also exist on double
sG kinks in the sector of topological degree 2. The attractive interaction
mediated by the fermions couples the solitons together, the corresponding spectral flow is not very different from the
similar results for a single sG kink in a model with shifted Yukawa interaction \cite{Brihaye:2008am} although the effects
of back-reactions become significant for the strong coupling, cf. similar observation for
the $\phi^4$ model \cite{Klimashonok:2019iya}.

\begin{figure}
 \begin{center}
\includegraphics[width=0.235\textwidth, trim={49, 20, 90, 60}, clip = true]{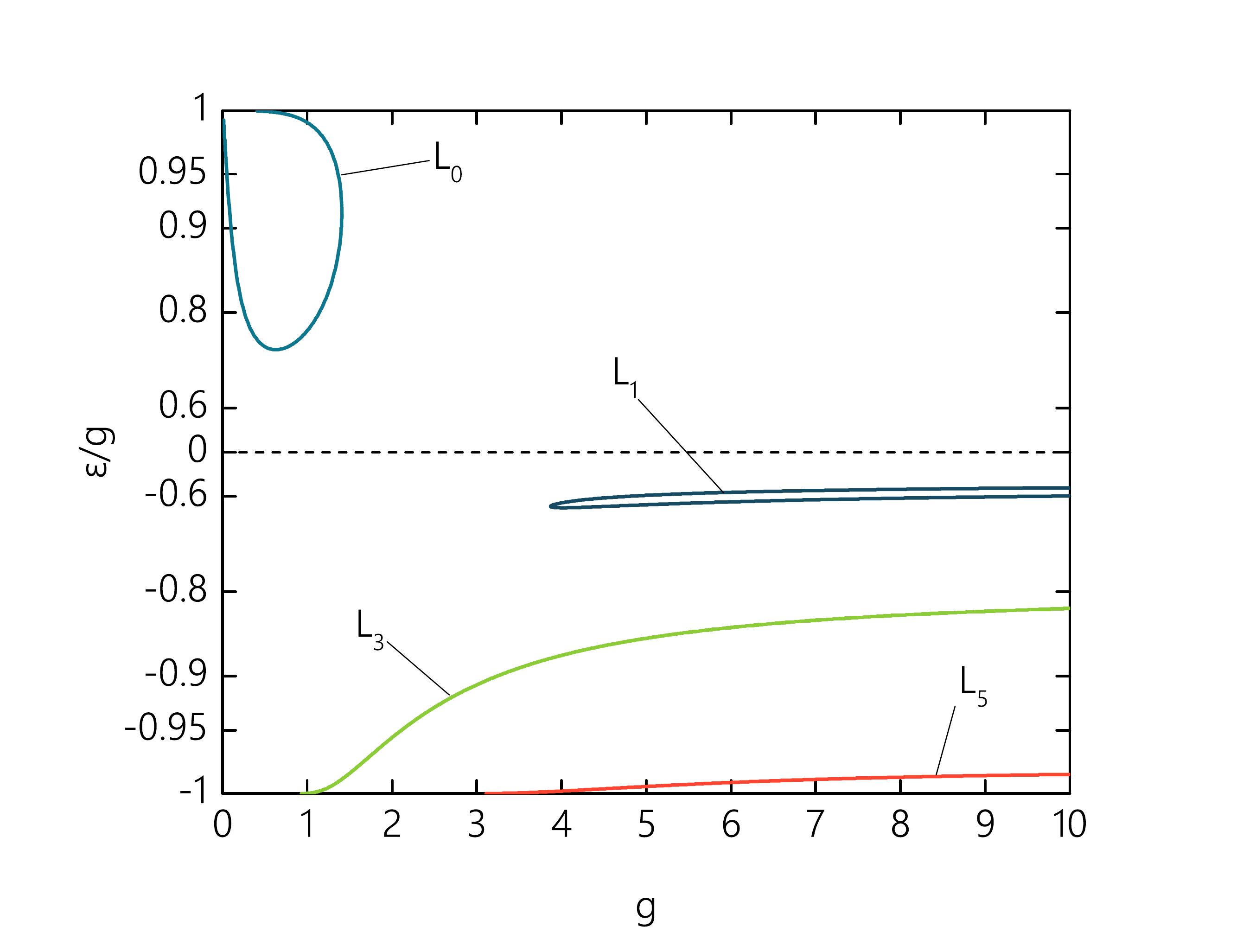}
\includegraphics[width=0.235\textwidth, trim={49, 20, 90, 60}, clip = true]{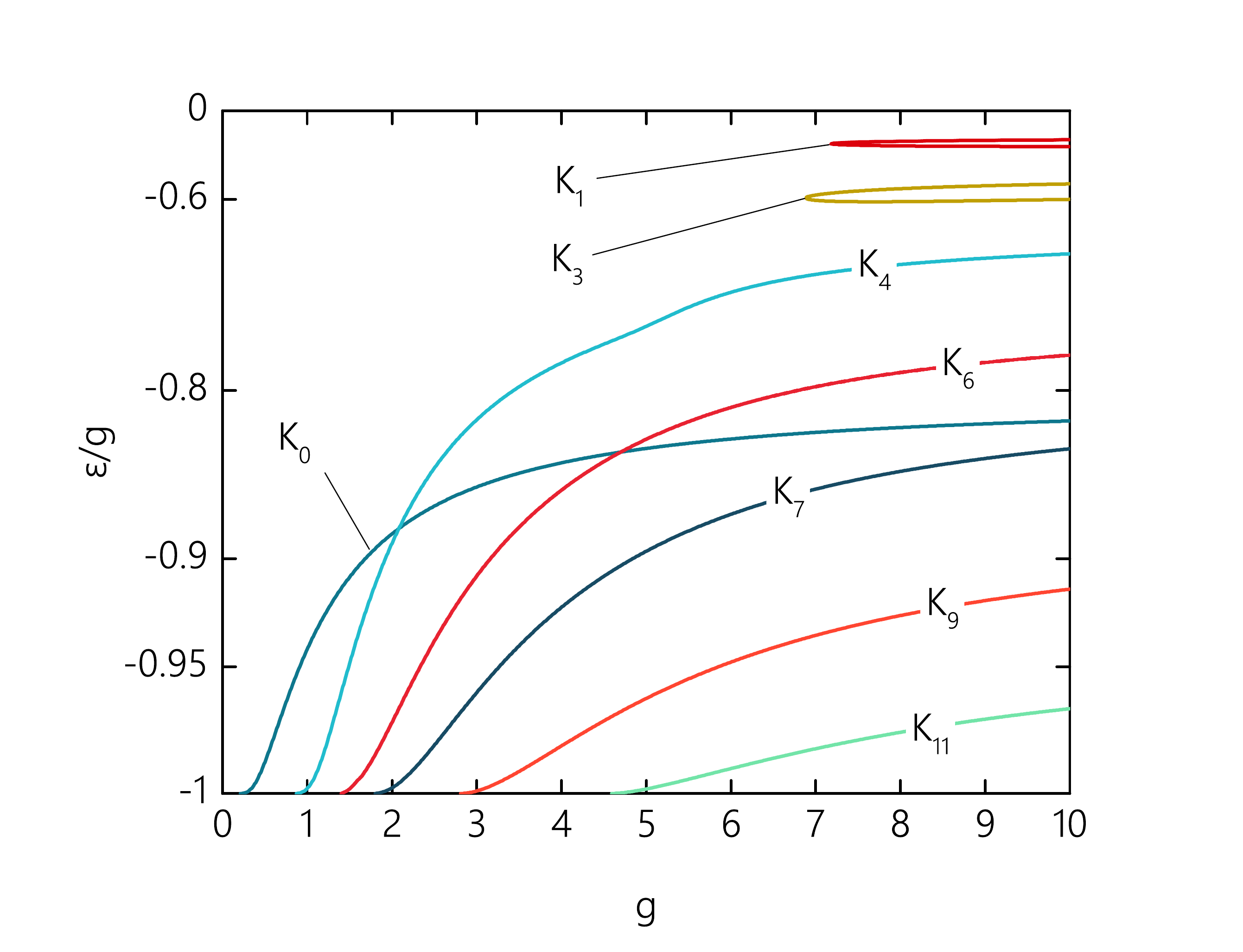}
\includegraphics[width=0.235\textwidth, trim={49, 20, 90, 60}, clip = true]{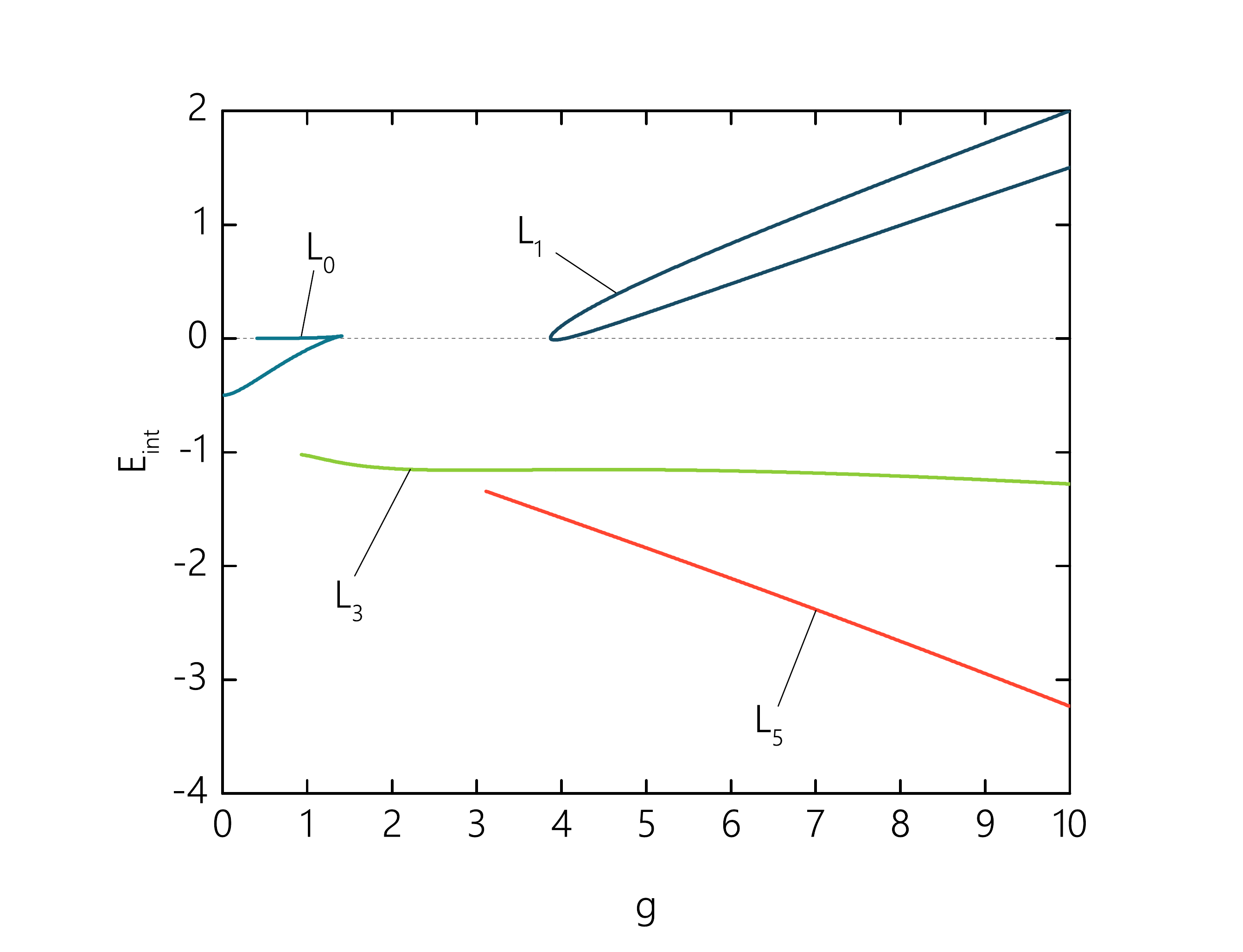}
\includegraphics[width=0.235\textwidth, trim={49, 20, 90, 60}, clip = true]{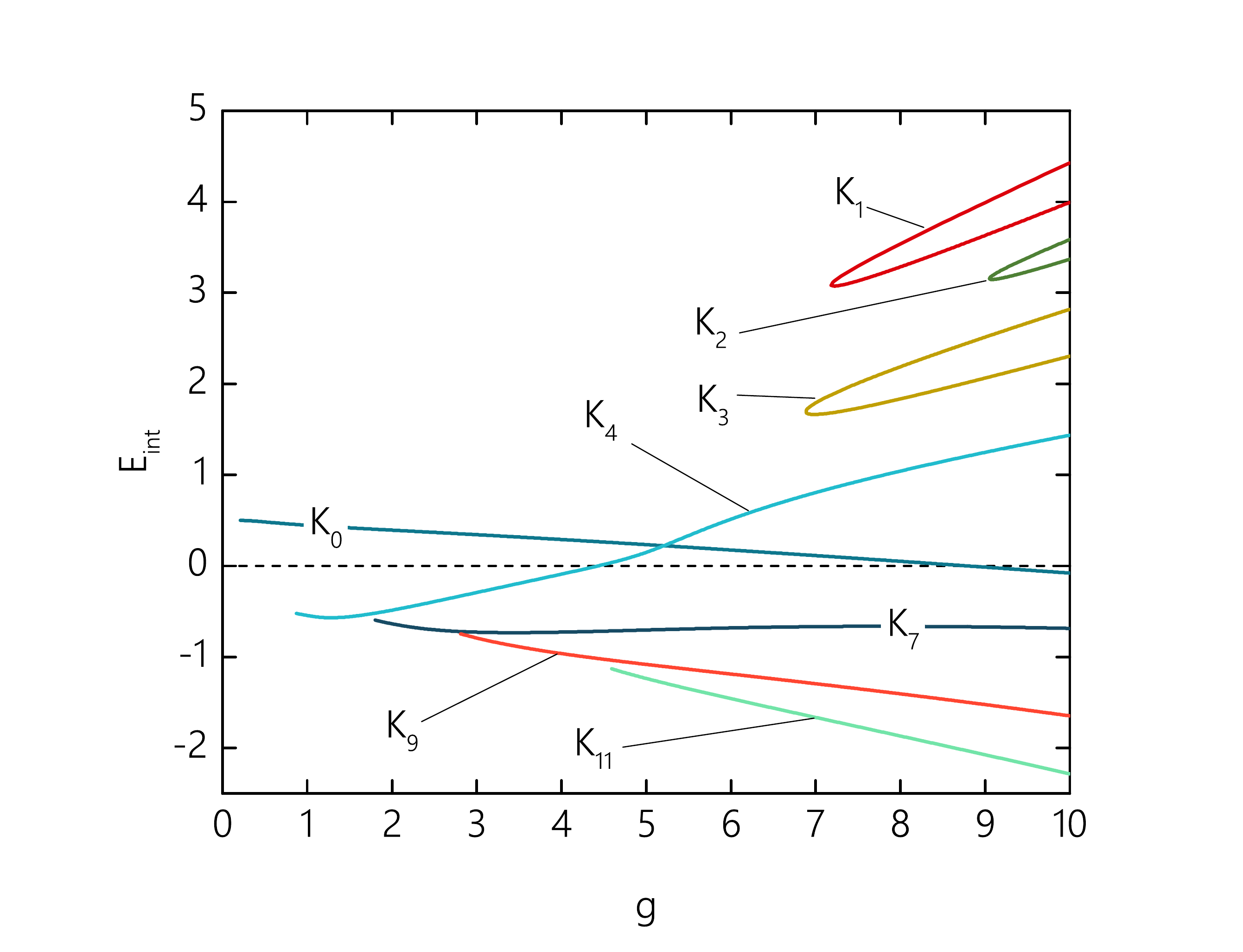}
\end{center}
 \caption{Fermion states bounded on the  $\phi^6$ kink-antikink pairs:
 normalized energy  ${\epsilon}/{g}$ of the localized modes and the interaction energy
 $E_\mathrm{int}$ as  functions of the Yukawa coupling $g$.}
\lbfig{Fig3}
\end{figure}

Now consider localized fermions in the $\phi^6$ model. For different multikink configurations we found a variety of
collective localized modes. First, by analogy with consideration above, we
consider kink-antikink pairs in the topologically trivial sector. Notably, there are
two distinct classes of vacua, they correspond to the $K\bar K$, or $(0,1)+(1,0)$ pair, and
$\bar K K$ pair $(1,0)+(0,1)$ \cite{Dorey:2011yw}. Considering the latter configuration, we found two
branches of localized fermions linked to the positive continuum,
these modes closely resemble similar nodeless collective modes $L_0$ on the $K\bar K $ pair in the sG model.
Higher modes, like $L_3$, also appear in the spectrum at large values of the coupling, see
Fig.~\ref{Fig3}. Another analogy with the collective modes $L_1$ localized on the $K \bar K  $ pair in the sG model
is that in the $\phi^6$ theory this mode is also disconnected from the continuum.
In a contrary, a collective potential for the fermions in the
presence if the $K\bar K$ pair has a raised central plateau, it does not support fermion bound states\footnote{
Interestingly, there is a certain similarity with the spectrum of linear scalar perturbations in the $\phi^6$ model
\cite{Dorey:2011yw}.}.

\begin{figure}
 \begin{center}
\includegraphics[width=0.19\textwidth, trim={49, 20, 100, 60}, clip = true]{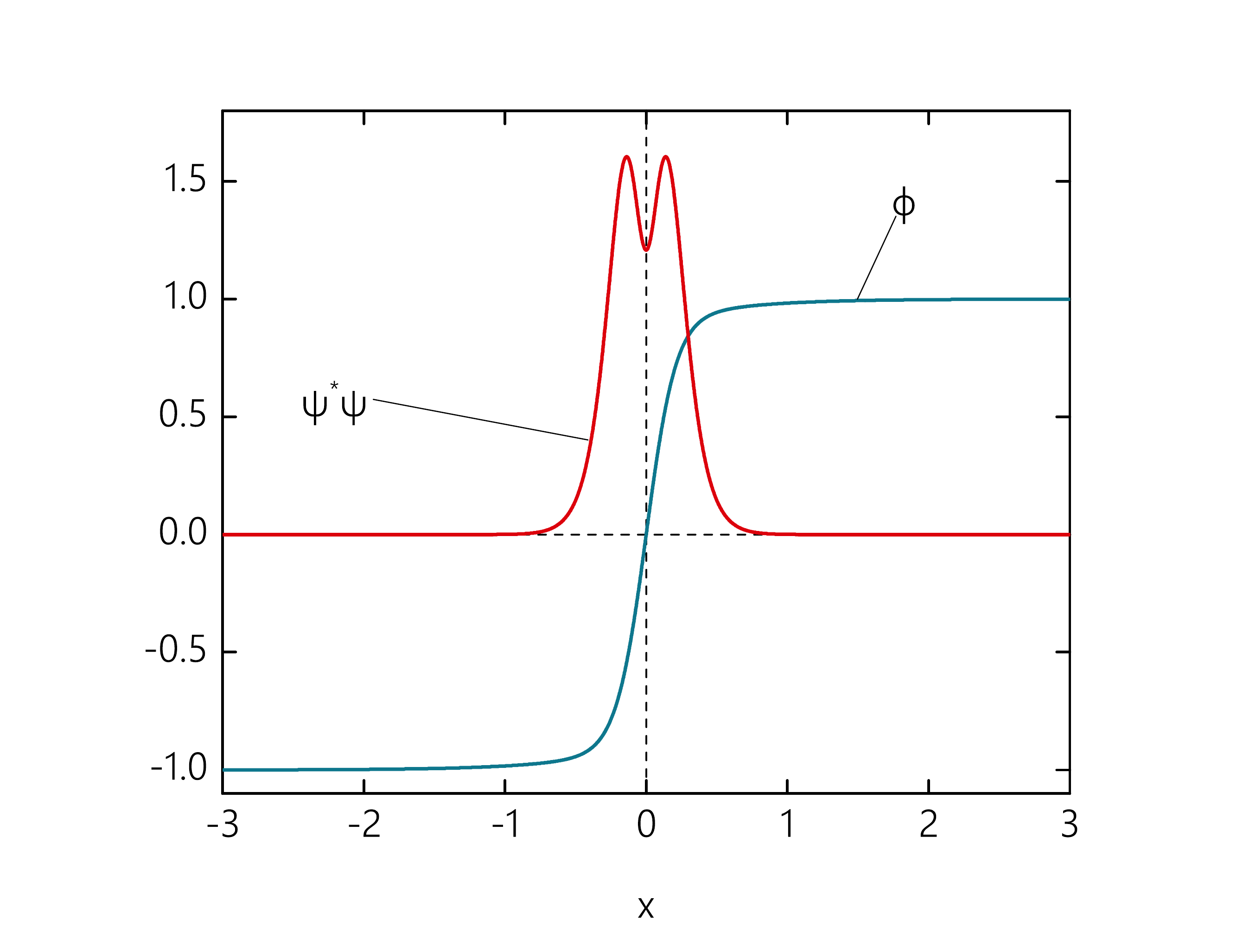}
\includegraphics[width=0.285\textwidth, trim={0, -30, 0, 0}]{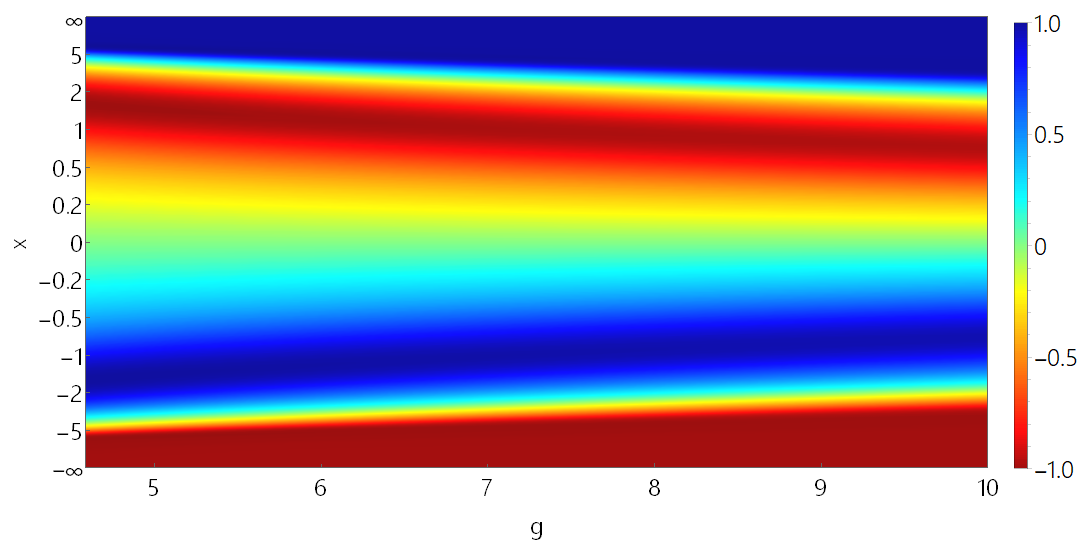}
\includegraphics[width=0.19\textwidth, trim={49, 20, 100, 60}, clip = true]{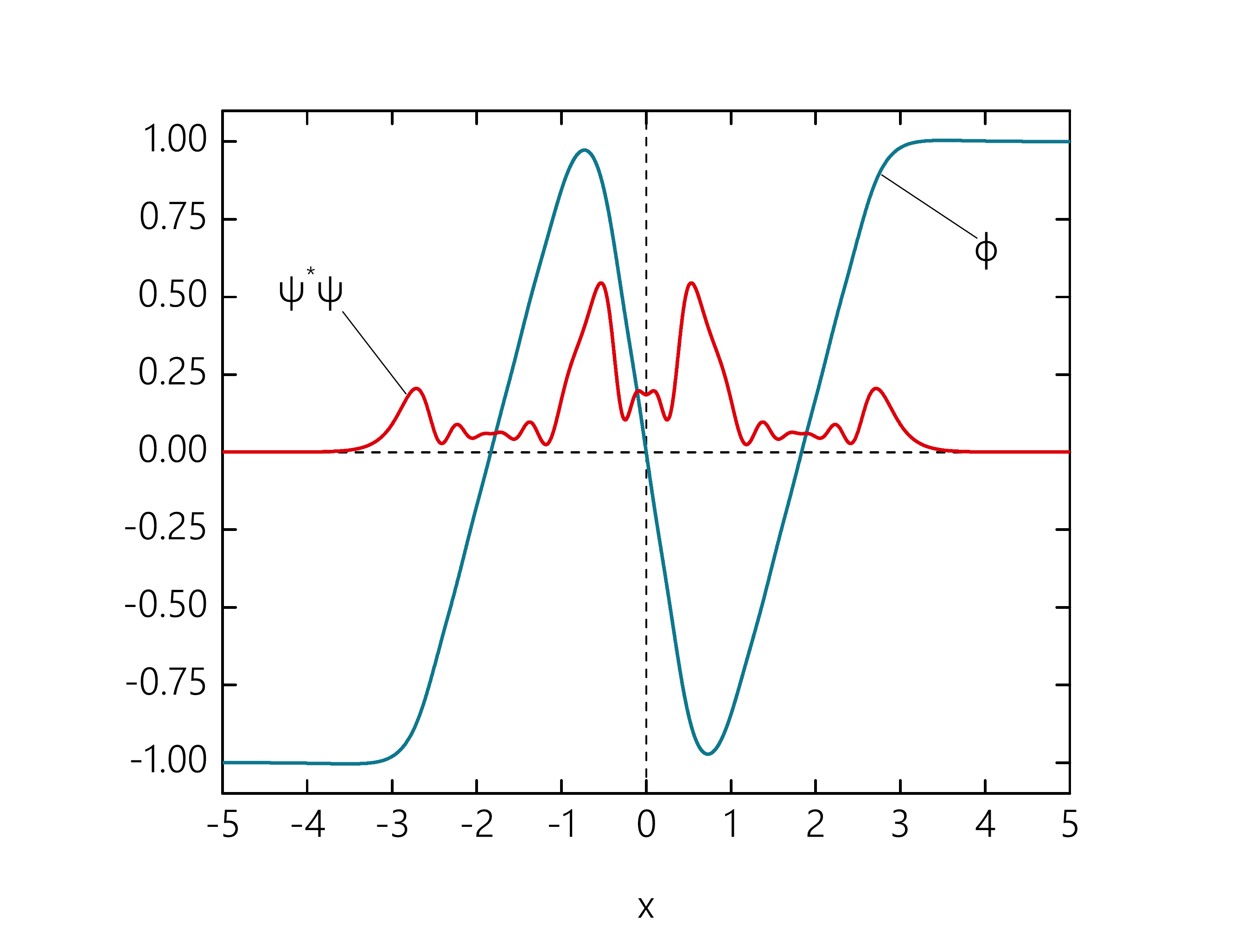}
\includegraphics[width=0.285\textwidth, trim={0, -30, 0, 0}]{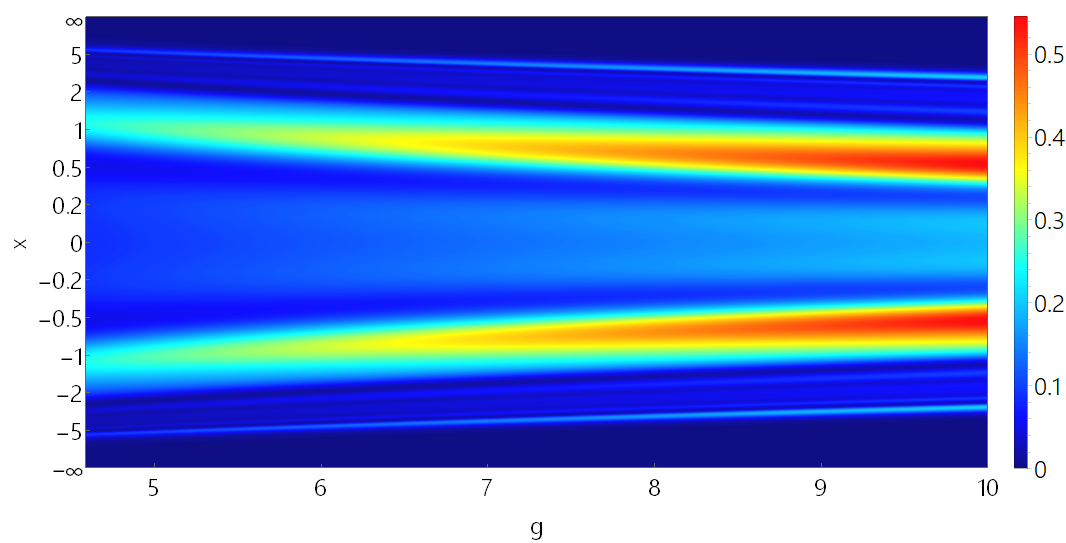}
\end{center}
 \caption{$\phi^6$ multikink configurations bounded by fermions. Profiles of the scalar field and
 fermion density distribution of the collective mode $K_0$  at $g=1$ (upper left) and the chain mode $K_{11}$ at
 $g=7$ (upper right), the scalar field of the chain of the kinks $(-1,1)+(1,-1)+(-1,1)$ bounded to the mode $K_{11}$
(bottom left) and fermionic density
distribution of this configuration (bottom right) as  functions of the Yukawa coupling $g$.}
\lbfig{Fig4}
\end{figure}
We found another family of solutions $K_n$, it represents two-kink configurations $(-1,0)+(0,1)$
bounded by collective fermion modes with a minimal node number $n$. Fig~\ref{Fig4} shows examples of such
states. More fermion bound states do appear for $g>1$, in particular the modes $K_1$ and $K_3$ do not emerge
from the continuum but appear at some bifurcation point. On the contrary, the modes
$K_2$, $K_4$ and other are linked to the negative energy continuum, see Fig~\ref{Fig3}.

A novel feature of the $\phi^6$ system with back-reaction is that it supports even more
complicated bounded multisoliton configuration which
represent kink-antikink chains with localized fermion modes. As an example, Fig~\ref{Fig4} exhibits
the chain of the kinks $(-1,1)+(1,-1)+(-1,1)$ bounded to the mode $K_{11}$.

The stability of the solutions can be tested by evaluation of the binding energy of the configurations,
which we define as a difference between the energy of the scalar components without fermions $E_0$ together
with the continuum threshold energy $g$,
and the total energy of the system: $E_{int}=E_0 +g-|\epsilon|-E_{\phi}$.
The results of our calculations are displayed in Fig.~\ref{Fig3}, bottom row. Clearly, some
configurations may become unstable with increase of the coupling $g$.

{\it Conclusions.~~}
In summary, we have shown that
localization of the backreacting fermion modes on multi-solitons
gives rise to a new mechanism of interaction enabling bound multisoliton
solutions to occur. More generally, we developed a new method of construction self-consistent solutions of
the coupled system of integral-differential equations for a multi-soliton configuration
coupled to fermions with back-reaction. This method is a powerful tool to study various
systems in a wide class of physical systems.
There might be a plethora of other applications including fermions on domain walls,
cosmic strings or vortices in the Abelian Higgs model, and monopole catalysis. We
hope to be addressing these questions in the near future.

\end{document}